\documentstyle[mprocl]{article}

\input{psfig}
\def\be{\begin{equation}}
\def\ee{\end{equation}}
\def\bea{\begin{eqnarray}}
\def\eea{\end{eqnarray}}

\begin{document}

\title{GAMMA-RAY BURSTS FROM NEUTRON STAR BINARIES}

\author{ G.J. MATHEWS}

\address{University of Notre Dame, Department of Physics, 
Notre Dame, IN 46556, USA}

\author{ J.R. WILSON, J. SALMONSON}

\address{Lawrence Livermore National Laboratory, Livermore, CA 94550}


\maketitle\abstracts{
We  report on  general relativistic hydrodynamic studies
which indicate several new physical processes which may contribute to
powering gamma-ray bursts in neutron star
binaries.    Relativistically
driven compression, heating, and collapse
of the individual stars can occur many seconds
before inspiral and merger.  This compression may produce a
neutrino burst of  $\sim 10^{53}$ ergs lasting 
several seconds.  The associated thermal neutrino emission
produces an $e^+-e^-$ pair plasma by $\nu \bar \nu$ annihilation.
We show first results of a simulated  burst which produces
$\sim 10^{51}$ erg in $\gamma$-rays.
We also discuss a preliminary study of the evolution of
the magnetic field lines attached to the fluid as the stars orbit.
We show that the relativistically driven
 fluid motion might lead to the formation
of extremely strong magnetic fields ($\sim 10^{17}$ gauss)
in and around the stars which could affect to the formation
and evolution of a gamma-ray burst.
}

It has been speculated for some time that inspiraling neutron stars
could provide a power source for cosmological gamma-ray bursts.
The rate of neutron star mergers (when integrated over the number of
galaxies out to high redshift)
could account for the observed GRB  event rate.
 The possibility that at least some $\gamma$-ray bursts
involve x-ray,  optical, or radio counterparts and are of cosmological origin has
recently received observational impetus.\cite{optical}
Mg I absorption and [O II] emission lines 
along the line of sight from the GRB970508 optical transient 
imply \cite{Metzger} that at least one burst is at a redshift
$Z \ge  0.835$.
If the bursts are cosmological, however,
they must entail a release
of $\sim 10^{51}$ erg in $\gamma$-rays on a
time  scale $\sim$ seconds.
Here we show preliminary calculations of a relativistically
driven GRB which is consistent with all of these constraints.

Previous, Newtonian and post Newtonian studies 
\cite{Janka} of the direct
merger of two neutron stars have found 
that the neutrino emission time scales are so short that
it would be difficult to drive a gamma-ray burst from this source.
However, our numerical studies
of the strong field relativistic hydrodynamics
of close neutron star binaries  in three spatial dimensions
\cite{wm95,wmm96,mw97}
have shown that neutron stars in a  close binary
can experience relativistic compression and heating
over a period of seconds.  This effect can even cause the stars to collapse
to two black holes prior to merger.
During the compression phase as much as 10$^{53}$ ergs in neutrinos can
be emitted  before the stars collapse.\cite{mw97}
This effect may provide a new mechanism to power cosmological
gamma-ray bursts and their x-ray and optical counterparts.
Here, we report on preliminary  efforts to better quantify this
release of neutrino energy around the binary and numerically
explore its consequences for the development of a $e^+-e^-$ plasma 
and associated  GRB.
 
In previous work \cite{mw97} we computed properties of
equal-mass neutron star binaries as a function of mass and EOS. 
From these we deduce that
compression, heating and collapse can occur at times
from a few seconds to a few hours before binary merger.
Our calculation of the rates of released binding energy
and neutron star cooling suggest that interior
temperatures as hot as 70 MeV are possible.  This leads to
several seconds of high neutrino luminosity, $L_\nu \sim 10^{53}$ 
erg sec$^{-1}$.
This much neutrino luminosity would convert to an $e^+-e^-$
pair plasma above the stars as  is also observed in supernova 
simulations.\cite{mayle} 
This plasma is a viable candidate source for
cosmological gamma-ray bursts.
 
We have begun to study
the transport of this neutrino flux above the neutron star
using a modified supernova code.\cite{mayle}  We find
entropies as high as $S/k\sim  10^{10}$  (i.e. few baryons)
in the pair plasma above the stars.
We have also made a preliminary calculation of the 
hydrodynamic evolution of the
pair plasma based upon our calculated neutrino emission
and an efficiency  (1-10\%) for the conversion of
neutrinos to $e^+-e^-$ pairs.  The results are quite encouraging.
We inject the pair plasma into a spherical grid at a rate consistent 
with the compression-induced thermal neutrino emission 
which itself is determined by
the gravitational wave emission time.\cite{mw97}
The plasma is evolved hydrodynamically until it  becomes optically
thin.  The calculation is then stopped and the escape of $\gamma$-rays
is calculated.  By this time the average temperature is about 10 eV, but the special
relativistic gamma-factor is $\approx 3\times 10^4$.    This produces an
 integrated  photon energy spectrum which is quite typical of observed bursts. 
It peaks at around 200 keV and extends to a few MeV.  Nearly all
energy deposited into $e^+-e^-$ pairs at the star surface ends up as $\gamma$ rays.
 
\begin{figure}
\psfig{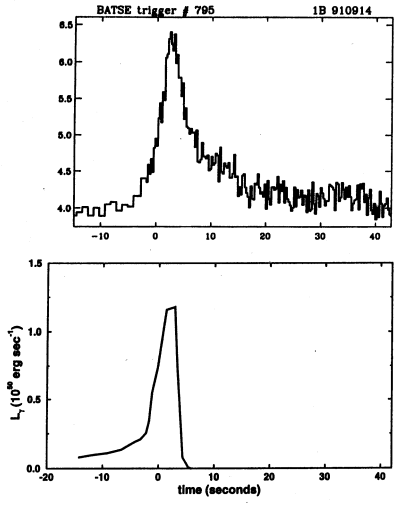}
\caption{Calculated gamma-ray burst luminosity (lower curve) compared
 with a similar single burst from the BATSE catalog.  The total released
energy from the calculated burst is $\sim 10^{51}$ erg.
\label{fig1}}
\end{figure}
 
Figure \ref{fig1}  shows a calculation of $\gamma$-ray burst 
luminosity as a function of time
compared with a typical "single-burst" from the BATSE catalog.  The
integrated energy in gamma-rays from the calculated burst is $\sim 10^{51}$ erg.
The similarity is remarkable 
 considering that there has been no parameter fitting in these
calculations.  Single-burst durations in the model vary from
$\sim 1$ to $10$ sec.  We also find that if the masses of the stars differ
by more than $\sim 5\%$ that the $\gamma$-ray emission
separates into two bursts spaced
a number of seconds apart.  Indeed, there are a number of BATSE bursts
indicating this morphology as well.  It is not clear, however, how
this compression scenario could lead to the typical multiple
peak structure observed in many bursts, without recourse
to a multiple shock mechanism.
 
In this regard we note \cite{wm98} the development of a toroidal fluid vorticity
in the neutron star interiors as a lowest stable configuration
for the binary.  This vorticity  is driven by relativistic force terms
arising from the motion of the stars with respect
to the curved three geometry.  Because
the circulation driving terms are purely relativistic, they
did not appear in previous
Newtonian or post-Newtonian calculations.
Nevertheless, they aid in the transport of neutrinos from the
neutron star interior and help
couple the stellar compression to internal heating by producing shocks.

An interesting possible consequence of this toroidal vorticity
is that it may lead to the development of extremely strong
magnetic fields within the stars.  If
the field lines are attached to the fluid vorticity, then 
the shearing of field lines could cause the magnetic
fields to grow until magnetic braking becomes important
near the equipartition limit,
$(H^2 / 8 \pi )  =  (\rho v^2 / 2)$.
For $\rho \sim 10^{15}$ g cm$^{-3}$ and $v \sim 0.1~c$.
This could imply a magnetic field as large as $H \sim 10^{17}$
gauss.

We have simulated the growth rate of the magnetic field by
introducing an electromagnetic  vector potential.
We followed the evolution
of the vector potential  for 10 msec (about one orbit)
assuming that the fluid is a perfect conductor.
The magnetic field energy 
exponentiated with an e-folding time of about 1 msec.
Thus, the field could build up very quickly to a magnitude
such that reconnection and back reaction of the fluid inhibits
further growth.  Due to the high density of fluid kinetic
energy in the vorticities, the limiting fields could approach
the equipartition limit. As the field grows it should bubble
from the surface (W. Kluzniak, Priv. Comm.).
Interactions of the bubbling field with the surface  pair plasma
might lead to the multiple peak structure observed in many
GRB's.

Work at University of Notre Dame
supported in part by DOE Nuclear Theory DE-FG02-95ER40934,
NSF PHY-97-22086, and by NASA CGRO NAG5-3818.
Work performed in part under the auspices
of the U.~S.~Department of Energy
by the Lawrence Livermore National Laboratory under contract
W-7405-ENG-48 and NSF grant PHY-9401636.


\section*{References}
\begin{thebibliography}{99}
 
\bibitem{optical}P. J.  Groot et al. 1997,  IAU Circ. No. 6584;
J. van Paradijs et al., 1997, Nature, 386,686.
 
\bibitem{Metzger}M.R. Metzger et al.,  IAU Circ. No. 6655 (1997).

\bibitem{Janka} H. T.  Janka \&  M. Ruffert 1996, A\&A, 307, L33 1996.

\bibitem{wm95}J.R. Wilson and G.J. Mathews, Phys. Rev. Lett., {\bf 75}, 4161 (1995).
 
\bibitem{wmm96}J.R. Wilson, G.J. Mathews, \& P. Marronetti, Phys. Rev. {\bf D54}, 131
7 (1996).
 
\bibitem{mw97}G. J. Mathews and J. R. Wilson, Astrophys. J., 482, 929 (1997).
 
\bibitem{wm98}J.R. Wilson and G.J. Mathews, Phys. Rev. Lett., {\it
submitted} (1997).

\bibitem{mayle}J. R. Wilson and R. W. Mayle, {\it Physics Reports},
{\bf 227}, 97-111 (1993).

\end{thebibliography}
\end{document}